# Yu-Shiba-Rusinov bands in ferromagnetic superconducting diamond

Gufei Zhang,[1]*† Tomas Samuely,[2]*† Naoya Iwahara,[3,4]*† Jozef Kačmarčík,[2] Changan Wang,[5] Paul W. May,[6] Johanna K. Jochum,[7] Oleksandr Onufriienko,[2] Pavol Szabó,[2] Shengqiang Zhou,[5] Peter Samuely,[2] Victor V. Moshchalkov,[8] Liviu F. Chibotaru,[3] Horst-Günter Rubahn[1]

[1]NanoSYD, Mads Clausen Institute and DIAS Danish Institute for Advanced Study, University of Southern Denmark, Alsion 2, DK-6400 Sonderborg, Denmark.

[2]Centre of Low Temperature Physics, Institute of Experimental Physics, Slovak Academy of Sciences & Faculty of Science, P. J. Safarik University, Kosice, Slovakia.

[3]Theory of Nanomaterials Group, KU Leuven, Celestijnenlaan 200F, B-3001 Heverlee, Belgium.

[4]Department of Chemistry, National University of Singapore, Block S8 Level 3, 3 Science Drive 3, 117543, Singapore.

[5]Helmholtz-Zentrum Dresden-Rossendorf, Institute of Ion Beam Physics and Materials Research, Bautzner Landstr. 400, 01328 Dresden, Germany.

[6]School of Chemistry, University of Bristol, Bristol BS8 1TS, United Kingdom.

[7]Laboratory of Solid State Physics and Magnetism, KU Leuven, B-3001 Heverlee, Belgium.

[8]Institute for Nanoscale Physics and Chemistry, KU Leuven, B-3001 Heverlee, Belgium.

*e-mail: gufei@mci.sdu.dk, tomas.samuely@upjs.sk, naoya.iwahara@gmail.com

†These authors contributed equally to this work.

**Abstract**

The combination of different exotic properties in materials paves the way for the emergence of their new potential applications. An example is the recently found coexistence of the mutually antagonistic ferromagnetism and superconductivity in hydrogenated boron-doped diamond, which promises to be an attractive system with which to explore unconventional physics. Here, we show the emergence of Yu-Shiba-Rusinov (YSR) bands with a spatial extent of tens of nanometers in ferromagnetic superconducting diamond using scanning tunneling spectroscopy. We demonstrate theoretically how a two-dimensional (2D) spin lattice at the surface of a three-dimensional (3D) superconductor gives rise to the YSR bands, and how their density-of-states profile correlates with the spin lattice structure. The established strategy to realize new forms of the coexistence of ferromagnetism and superconductivity opens a way to engineer the unusual electronic states and also to design better performing superconducting devices.

## Introduction

The interplay between different coexisting physical properties induces emergent phenomena, opening a way to design new electronic devices. The investigated mutual interactions between electrical conductivity/superconductivity and magnetism give rise to, for example, colossal magnetoresistance in manganese oxides[1] and superconducting pairing by spin fluctuation in iron pnictides. These mutual interactions are also at the heart of Majorana bound states, which arise from the interaction of chiral magnetic fields with a superconducting condensate and are natural candidates for topologically protected qubits[2-5]. In this context, a new intriguing case where superconductivity and ferromagnetism coexist was recently found in hydrogenated boron-doped diamond films[6].

In order to assess the effect of magnetic perturbation on superconductivity, one powerful approach is the analysis of the influence of individual magnetic impurities or groups of impurities on the superconducting state. Their effect has been extensively revealed in various superconductors[7], where it was found that individual local magnetic moments interacting with superconducting electrons give rise to in-gap localized states, called Yu-Shiba-Rusinov (YSR) states[8-10]. For the investigation of such localized YSR states, scanning tunneling microscopy/spectroscopy (STM/S) is a methodology of choice, because it directly provides the in-gap local density of states (DOS) with high spatial and energy resolution[7,11-16].

Here, we performed STM/S measurements on ferromagnetic superconducting diamond. In contrast to conventional YSR states localized within few nanometers, we find YSR states with a spatial extent of tens of nanometers, indicating strong delocalization of the electronic excitations. We model the hydrogenated boron-doped diamond as a 3D superconductor with a 2D spin lattice at the surface. In order to reproduce the STM/S results within the Bogoliubov-de Gennes (BdG) theory, we assume that hydrogen atoms adsorbed at the surface of boron-doped diamond nanocrystallites arrange in islands of ordered 2D lattices, giving rise to correspondingly ordered islands of magnetic impurities. The resulting impurity states correspond to YSR bands, whose properties strongly depend on the structure of the underlying lattice of localized spins. Hence, systems such as hydrogenated boron-doped diamond are good platforms for hosting such YSR bands.

## Results

### Hydrogenated boron-doped diamond

Heavily boron-doped polycrystalline chemical vapor deposited (CVD) diamond films with a hydrogen-terminated surface were used for this research. Structural and compositional analyses indicate that the polycrystalline diamond films are metal-free[6], and have a thickness of ~900 nm, a mean grain size of ~800 nm, and a boron concentration of ~$1.5 \times 10^{21}$ cm$^{-3}$. Raman spectroscopy confirms that the sample surface is hydrogen terminated[6].

### Coexistent superconductivity and ferromagnetism

The onset of the superconducting state ($\rho = 0$) takes place at 2.7 K in the hydrogenated boron-doped diamond. To gain insight into the dimensionality of the system, we measured the resistive superconducting transition $\rho(T)$ in different magnetic fields, $\mu_0 H$, which were applied perpendicular and parallel to the sample (see Fig. 1A and 1B). In contrast to the linear dependences of the offset temperature of the resistive superconducting transition on the applied magnetic fields, the $\mu_0H_{C2}$-$T$ phase boundaries as determined by the onset temperature of the resistive superconducting transition, are in good agreement with quadratic fits (see Fig. 1C). Due to the quantum confinement of the superconducting order parameter in the presence of out-of-plane grain boundaries and twin boundaries, the $\mu_0H_{C2}$-$T$ phase boundary measured in perpendicular fields overshoots the in-plane phase boundary at finite temperatures, revealing a crystallite geometry-induced anomalous superconducting anisotropy in the diamond films. When extrapolating the $\mu_0H_{C2}$-$T$ phase boundaries down to zero-temperature, the quadratic fits converge at $\mu_0H_{C2}(0\text{ K}) \sim 4$ T (see Fig. 1C). According to $\xi_{GL} = [\Phi_0/2\pi \mu_0H_{C2}(0\text{ K})]^{0.5}$ with $\Phi_0$ being the flux quantum, the Ginzburg-Landau coherence length $\xi_{GL}$ is determined to be 9.1 nm. Despite the superconducting anisotropy at finite temperatures, the substantial distinction between $\xi_{GL}$ and the film thickness indicates the 3D nature of the superconductivity.

The magnetization hysteresis loops measured below the superconducting transition temperature show characteristics of both superconductivity and ferromagnetism (see Fig. 1D). The former is featured with the central peak and the low-field diamagnetism in the virgin curve, while the latter is reflected in the overall shape of the loops. The temperature-induced evolution of such anomalous hysteresis loops has been reported in Ref. 6. The presence of ferromagnetism in various forms of metal-free carbon has remained an intriguing question for decades. Although both $sp^2$/$sp^3$ defects and hydrogenation have been commonly employed to explain the ferromagnetism observed in different forms of carbon[6], polycrystalline diamond rich in $sp^2$/$sp^3$ defects but without hydrogenation did not show any ferromagnetic behavior[17]. Furthermore, theoretical calculations demonstrated that unhydrogenated carbon sites and hydrogen impurities can give rise to local magnetic moments in graphene and diamond, respectively[18-20].

The substantial distinction between the in-plane and out-of-plane magnetization loops indicates the in-plane easy axis of magnetization (see Fig. 1D). Taking into account the robust 3D nature of the superconductivity, the diamond films cannot be simply treated as conventional ferromagnets in the form of thin film, where similar magnetic anisotropy is mainly caused by shape anisotropy[21]. The anomalous superconducting anisotropy mentioned above cannot be a cause of the magnetic anisotropy, either, because at 1.8 K and in the magnetic field range of 0-0.4 T, the diamond films demonstrate no superconducting anisotropy (see Fig. 1C). Accordingly, the observed magnetic anisotropy is highly likely due to the anisotropic distribution of magnetic impurities.

**Yu-Shiba-Rusinov bands observed in direct local measurements**
As revealed by STM/S measurements performed at 0.5 K on the surface of the hydrogenated boron-doped diamond (see Fig. 2A), the zero-bias tunneling conductance (ZBC) map demonstrates strong modulations of the local DOS on the nanoscale, which appear as 'puddles' (see Fig. 2B and 2C). The tunneling conductance spectra acquired over the puddles show symmetric maxima and minima around the zero energy inside the superconducting gap 2Δ (see Fig. 3A), which are fingerprints of the YSR effect. Strikingly, in contrast to the localized YSR states which are restricted to a few atomic distances in 3D systems[14], these collective excitations extend up to tens of nanometers, revealing the delocalization of the YSR states and the emergence of YSR bands. The extraordinarily large spatial extension of our observed YSR bands can be also seen in the spatial dependence of the integrated in-gap DOS (see Fig. 3B).

The observed YSR bands demonstrate a variety of DOS profiles (see Fig. 3A, Fig. 4A and 4B). Apart from the pair of peaks at finite energies ($E_{YSR} \sim \pm 0.39\Delta$), the tunneling spectra S1 and S4 both show a third peak at zero energy inside the superconducting gap $2\Delta \sim 3$ meV. Along with the peaks at $E_{YSR} \sim \pm 0.22\Delta$ in Spectrum S2, a pair of kinks ($E_{YSR} \sim \pm 0.39\Delta$) emerge at the shoulders of the peaks. Spectrum S3 exhibits two peaks at $E_{YSR} \sim \pm 0.35\Delta$ and two kinks at $E_{YSR} \sim \pm 0.21\Delta$, while Spectrum S5 reveals the emergence of two pairs of peaks at $E_{YSR} \sim \pm 0.61\Delta$ and $E_{YSR} \sim \pm 0.2\Delta$, respectively. Note that among these YSR bands, even S5 shows a spatial extent of ~9 nm (see L4 in Fig. 3A). In contrast to these YSR bands, the tunneling spectra outside the puddles exhibit Bardeen-Cooper-Schrieffer-like gap structures (see Spectrum S6 in Fig. 4A and 4B).

**Theoretical modeling of Yu-Shiba-Rusinov bands with a tight-binding Bogoliubov-de Gennes approach**

In order to give insights into the origin of the various DOS profiles (see Fig. 4A and 4B), we developed a microscopic theory. The ferromagnetism can arise from both bulk hydrogen impurities[20,22,23] and hydrogen islands at the surface, and the mechanism is still largely unknown. Since we observed extended but finite (several tens of nm) regions with non-negligible in-gap DOS, we speculate that its origin is the presence of islands of magnetic impurities located at or close to the surface, and which interact with the bulk superconducting state. Moreover, although the present STM/S measurements do not have atomic resolution and the character of the distribution of the impurities cannot be determined (see Materials and Methods), it should have strong correlation with the type of the surface. The surface of the polycrystalline nanodiamond films is far from flat, and there are (111) and (100) surfaces[6], giving rise to triangular and rectangular spin lattices on them, respectively (see Fig. 5). Given the strong exchange interaction between the magnetic centers (with Curie temperature >400 K), the spins on the surface can be treated as classical ones.

The electronic bands in the presence of a spin lattice were calculated with a tight-binding BdG approach with the localized YSR states on spin sites as the basis[4,5]. The BdG equation in the presence of the classical spin $S$ at $\mathbf{R}_j$ is given by

$$\begin{pmatrix} \xi_k & \Delta \\ \Delta & -\xi_k \end{pmatrix} \mathbf{w_k} - \sum_j JS \begin{pmatrix} 1 & 0 \\ 0 & 1 \end{pmatrix} e^{-i\mathbf{k}\cdot\mathbf{R}_j} \mathbf{w}(\mathbf{R}_j) = E\mathbf{w_k}$$

where $\mathbf{w_k}$ stands for the quasi-particle states characterized by wave vector $\mathbf{k}$, $\mathbf{w}(\mathbf{R}_j)$ indicates the localized YSR state at site $j$, $E$ is energy, $\xi_k$ is the free electron energy with respect to chemical potential, $J$ is the exchange coupling parameter, $S = 1/2$ is spin, and $\Delta$ is the superconducting order parameter. We transformed this equation into the tight-binding form on the basis of localized YSR states, and then calculated the quasi-particle bands (see Methods). To reproduce the experimental DOS, we treated the dimensionless exchange interaction $\alpha = \pi v_0 JS$ between the spins and conduction electrons ($v_0$ is the DOS of the electronic band at the Fermi energy) and the distance between the spin sites as variables.

Among the experimentally obtained DOS, we focused upon S1, S2 and S3 which clearly show the band nature (Fig. 3A). Figure 4C shows the theoretical DOS calculated using rectangular lattices for S1 and S2, and a triangular lattice for S3 with $\alpha = 0.7$ (see Materials and Methods for other parameters). The dimensions of the rectangles are $9 \times 10$ and 9 × 13 for S1 and S2, respectively, and the length of the side of equilateral triangle is 12 for S3, all in units of diamond lattice constant, $a = 3.57$ Å. The numbers and the positions of the maxima of the DOS within the superconducting gap are in good agreement with the experimental data. Qualitatively different DOS profiles were obtained for S1 and S2 using the same spin lattice structure and $\alpha$, because the interaction between the YSR states oscillates with respect to the distance. The good agreement between the theoretical modeling and the experimental data confirms that the observed in-gap excitations are YSR bands. Moreover, the diameter of the YSR localized state[14] and the distance between the nearest lattice points are comparable to each other, which is consistent with the absence of the spatial variation of the DOS in our experimental spectra S1-3 obtained with the present spatial resolution of STM/S (see Fig. 3A). Despite the fact that the magnetic impurities tend to destroy the superconductivity, the superconducting temperature of the ferromagnetic diamond is almost unchanged, compared to other doped diamond superconductors[24,25]. This is explained by the low-concentration of the hydrogen-related magnetic impurities on surface as readily understood from the dimensions of the lattices used for the calculations. The theoretical analysis also suggests that the present system is close to the regime of gapless superconductivity due to the classical spins[9].

**Discussion**

In summary, we unveiled the presence of long-range coherent YSR bands in a 3D ferromagnetic superconductor, heavily boron-doped diamond with hydrogen-terminated surface. In contrast to localized YSR states, the observed YSR bands show a spatial extent up to tens of nanometers. Limited by the resolution of STM/S measurements on polycrystalline diamond, the spatial distribution of magnetic impurities cannot be extracted from our experiments. By developing a microscopic theory for a 3D superconductor with a 2D spin lattice at its surface, we reproduced the experimental DOS profiles of the observed YSR bands in ferromagnetic superconducting diamond reasonably well. Apart from the Fulde-Ferrell-Larkin-Ovchinnikov state[26,27] and the so-called domain wall superconductivity[28,29], introduction of the sparse magnetic impurity lattice is another form of the coexistence of the two mutually antagonistic orderings, ferromagnetism and superconductivity. The present results suggest the possibility to construct hybrid systems consisting of, *e.g.*, sparse 2D ferromagnetic lattice and superconductor with spin-orbit coupling[5] to realize unconventional topological superconducting phases.

**Materials and Methods**

**Sample preparation**

The heavily boron-doped and hydrogenated polycrystalline diamond films were grown using hot filament chemical vapor deposition (CVD). The substrates were undoped single-crystal Si (100) with a 30 nm-thick layer of $SiO_2$ on the surface. The wafers were seeded with a colloidal suspension of diamond nanoparticles (diameter ~20 nm) using an electrospray process[30] (seeding density ~$3 \times 10^{10}$ cm$^{-2}$). The seeded substrate was then placed 3 mm below a 2200 °C tantalum filament in the CVD reactor. A gas mixture of 0.6% $CH_4$ in $H_2$ (total flow 200 sccm) was thermally dissociated using the hot filament for diamond growth onto the substrate, which was maintained at 800 °C. Boron doping was realized by adding diborane ($B_2H_6$) to the gas mixture with a $B_2H_6$/$CH_4$ ratio of 5%. In order to ensure that the diamond surface was hydrogen terminated, after deposition was complete the $CH_4$ and $B_2H_6$ flows were switched off while the sample remained under the filaments for 1 min in pure hydrogen gas. The filaments were then switched off, allowing the sample to cool back to room temperature under flowing $H_2$ in ~30 min[6].

**Electrical transport and magnetization measurements**
The electrical transport properties and magnetization of the samples were measured using a Heliox $^3$He cryostat and a MPMS3 system, respectively.

**Scanning tunneling microscopy/spectroscopy**
The differential tunneling conductance spectra were acquired by means of a home-made scanning tunneling microscope[31] (STM) cooled to 0.5 K by a commercial Janis SSV cryomagnetic system with a $^3$He refrigerator. The atomically sharp gold STM tip was obtained *in situ* by systematically impaling the tip into a clean gold surface. To measure the tunneling current spectra, the feedback loop was turned off and the tunneling current at different bias voltages applied to the tip was recorded. The applied bias voltage was swept from -10 to +10 mV with an interval of ~20 µV, which is smaller than the width of the in-gap peaks by one order of magnitude. The initial tunneling resistance was 5 MΩ. Since the gold tip features a constant density of states, the differential conductance derived numerically from the current spectra represents the local density of states of the diamond surface. Despite the atomic sharpness of the STM tip, the spatial resolution of our STM/S measurements was limited to ~3 nm, due to the surface roughness of the lab-grown polycrystalline diamond films. As a result of the newly formed nucleation sites, the surface of the diamond crystallites is corrugated on the nanoscale and far from flat. The nanoscale corrugations give rise to additional noise, which limits the spatial resolution of the STM/S measurements.

**Theoretical modeling**
The Bogoliubov-de Gennes equation was transformed into the tight-binding form on the basis of localized YSR states[4]

$$\mathbf{w}(\mathbf{R}_i) = \sum_j -JS \begin{pmatrix} EI_0(R_{ij}) + I_1(R_{ij}) & \Delta I_0(R_{ij}) \\ \Delta I_0(R_{ij}) & EI_0(R_{ij}) - I_1(R_{ij}) \end{pmatrix} \mathbf{w}(\mathbf{R}_j) \quad (1)$$

where $R_{ij} = |\mathbf{R}_i - \mathbf{R}_j|$, $I_0(R) = (2\pi)^{-3}\int d^3\mathbf{k}\, e^{i\mathbf{k}\cdot\mathbf{R}}/(E^2 - \xi_k^2 - \Delta^2)$ and $I_1(R) = (2\pi)^{-3}\int d^3\mathbf{k}\, \xi_k e^{i\mathbf{k}\cdot\mathbf{R}}/(E^2 - \xi_k^2 - \Delta^2)$. These integrals were evaluated under the condition of $(\hbar k_F)^2/2m_e \gg \hbar\omega_D \gg \Delta > E$ as

$$I_0(R) \simeq -\frac{\pi\nu_0}{\sqrt{\Delta^2 - E^2}} \frac{\sin(k_F R)}{k_F R} \exp\left(-\frac{\sqrt{\Delta^2 - E^2}}{\hbar\upsilon_F} R\right),$$

$$I_1(R) \simeq -\frac{\pi\nu_0}{k_F R}\left\{\left[\exp\left(-\frac{\sqrt{\Delta^2 - E^2}}{\hbar\upsilon_F}R\right) - 1\right]\cos(k_F R) + \frac{\sqrt{\Delta^2 - E^2}}{\hbar\upsilon_F k_F}\exp\left(-\frac{\sqrt{\Delta^2 - E^2}}{\hbar\upsilon_F}R\right)\sin(k_F R)\right\}$$

$$+ \frac{\pi\nu_0}{k_F R}\frac{2}{\pi}\left[-\cos(k_F R)\text{Si}\left(\frac{\omega_D R}{\upsilon_F}\right) - \sin(k_F R)\frac{\sin\left(\frac{\omega_D R}{\upsilon_F}\right)}{k_F R}\right] \quad (2)$$

where $\hbar$ is the reduced Planck constant, $k_F$ and $v_F$ are the wave vector and velocity at the Fermi energy, respectively, $m_e$ is the mass of electron, $\omega_D$ is the Debye frequency, and $\mathrm{Si}(x)$ is the sine integral. The electron attraction and $\Delta$ are assumed to be finite when $|\xi_k| < \omega_D$. The expression of $I_1$ was improved so that small distances can be accurately treated in comparison with the previous work based on asymptotic treatment[4].

When there is only one impurity, the eigenvalues $E$ of Eq. (1) reduce to[8-10]:

$$E = \pm \Delta \frac{1-\alpha^2}{1+\alpha^2}$$

where $\alpha = \pi v_0 JS$. The energy levels appear within the superconducting gap.

Assuming the periodicity of the impurity centres, Eq. (1) was Fourier transformed with the period of impurity centres

$$\lambda_\mathbf{q} \mathbf{w}'_\mathbf{q} = \begin{pmatrix} \frac{\sqrt{1-\lambda_\mathbf{q}^2}}{\alpha} & -1 \\ -1 & \frac{\sqrt{1-\lambda_\mathbf{q}^2}}{\alpha} \end{pmatrix} \mathbf{w}'_\mathbf{q} - \begin{pmatrix} \lambda_\mathbf{q} \tilde{I}_{0,\mathbf{q}} + \tilde{I}_{1,\mathbf{q}} & \tilde{I}_{0,\mathbf{q}} \\ \tilde{I}_{0,\mathbf{q}} & \lambda_\mathbf{q} \tilde{I}_{0,\mathbf{q}} - \tilde{I}_{1,\mathbf{q}} \end{pmatrix} \mathbf{w}'_\mathbf{q}$$

(3)

Here, $\mathbf{w}'_\mathbf{q} = \sum_j \mathbf{w}(\mathbf{R}_j) \exp(-i\mathbf{q}\cdot\mathbf{R}_j)$ is the Fourier transformation of $\mathbf{w}(\mathbf{R}_i)$, $\lambda_\mathbf{q} = E/\Delta$, $\tilde{I}_{n,\mathbf{q}}$ ($n = 0,1$) is defined by

$$\tilde{I}_{n,\mathbf{q}} = \sum_{j(\neq 0)} e^{-i\mathbf{q}\cdot\mathbf{R}_j} \tilde{I}_n(R_j)$$

(4)

$\mathbf{R}_0 = 0$, $\tilde{I}_0(R) = I_0(R)/I_0(0)$, and $\tilde{I}_1(R) = I_1(R)/[\Delta I_0(0)]$. Finally, by solving the eigenvalue problem (3), we obtained a nonlinear equation for each wave vector $\mathbf{q}$

$$\lambda_\mathbf{q} = \frac{\sqrt{1-\lambda_\mathbf{q}^2}}{\alpha} - \lambda_\mathbf{q} \tilde{I}_{0,\mathbf{q}} \pm \sqrt{(1+\tilde{I}_{0,\mathbf{q}})^2 + \tilde{I}_{1,\mathbf{q}}^2}$$

(5)

Neither spin-orbit coupling nor edge of the lattice is considered in our theory, and hence, the scenario of the emergence of exotic edge states discussed in Ref. 5 does not apply.

By numerically solving this equation, we obtained the quasi-particle energy band. For the simulations, $k_F \approx 1\text{-}3 \times 10^9$ m$^{-1}$ and $v_F \approx 7\text{-}8$ eV Å were taken from Ref. 32, and $a = 3.57$ Å and $\omega_D \approx 160$ meV for diamond were taken from Ref. 33. With these parameters, $k_F a \approx 1$, $\hbar v_F/a \approx 2$ eV, $\omega_D/(v_F/a) \approx 0.08$, and $\Delta/(\hbar v_F k_F) \approx 0.0005$. The shape of the impurity lattice is either a rectangle or regular triangle. The DOS was convoluted with Gaussian function with standard deviation $\sigma/\Delta = 0.14, 0.10, 0.09$ for S1, S2 and S3, respectively. The height of the DOS, which depends on the distance between the STM tip and the surface[14], was rescaled to match the experimental DOS.

**Acknowledgments**

**Funding:** G.Z. thanks the Fabrikant Mads Clausens Fond. T.S., J.K., O.O., P.Sz. and P.S. are supported by APVV-18-0358, VEGA 1/0743/19, VEGA 2/0149/16, COST action CA16218 Nanocohybri, and H2020 Infraia 824109 European Microkelvin Platform. N.I. is supported by GOA from KU Leuven and the scientific research grant R-143-000-A80-114 of the National University of Singapore. J.K.J. thanks the Hercules foundation.

**Author contributions:** G.Z. conceived the study and designed the experiments. P.W.M. prepared the samples. T.S., O.O. and P.S. performed the scanning tunneling spectroscopy measurements. J.K., G.Z. and P.Sz. measured the electrical transport properties. S.Z., J.K.J. and C.W. carried out the SQUID measurements. G.Z. and T.S. analysed the data. N.I. and L.F.C. performed the theoretical modeling. G.Z., N.I. and L.F.C. prepared the manuscript with inputs from T.S. and P.W.M. P.S., V.V.M. and H.G.R took part in the discussions and correction of the manuscript.

**Competing interests:** The authors declare no competing interests.

**Data and materials availability:** The data that support the findings of this study are available within the article. All the other data supporting the findings of this study within the article are available on request from the corresponding authors.

**Figures**

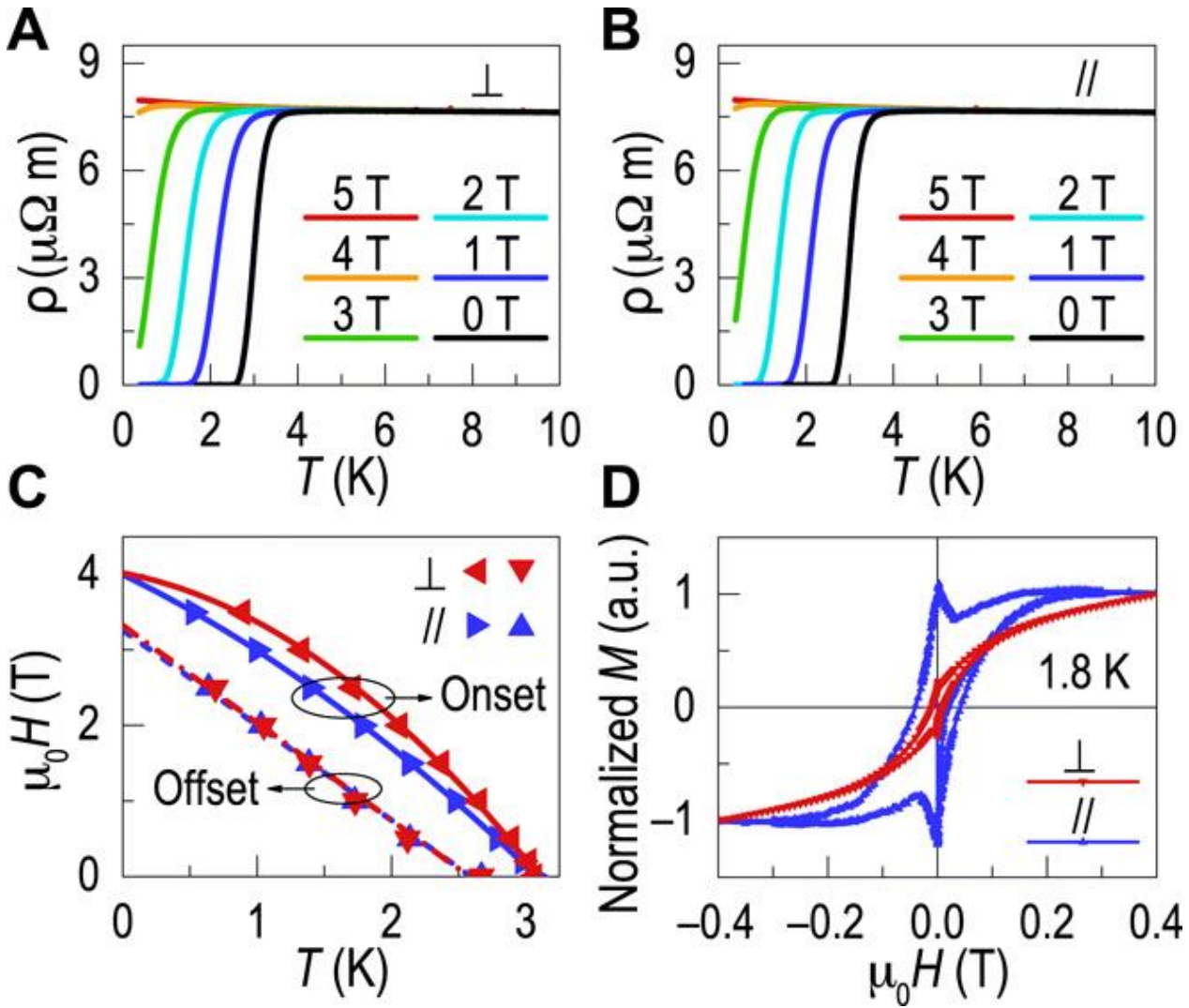

Fig. 1. **Coexistent superconductivity and ferromagnetism in hydrogenated boron-doped diamond.** Resistive superconducting transitions in (**A**) perpendicular ($\perp$) and (**B**) parallel (//) applied magnetic fields, respectively. (**C**) The temperature dependences of the critical field, $\mu_0H(T)$, measured in magnetic fields perpendicular and parallel to the sample. To build up the $\mu_0H$–$T$ phase boundaries, a criterion was set at 95% of the normal-state resistivity to determine the onset critical temperature, and the resistive superconducting transition was linearly extrapolated down to $\rho = 0$ for the determination of the offset critical temperature in different magnetic fields. The $\mu_0H$–$T$ phase boundaries are extrapolated to zero temperature by quadratic fits (solid curves) and linear fits (dashed lines), respectively. (**D**) Magnetization hysteresis loops measured at 1.8 K in magnetic fields perpendicular and parallel to the sample. The magnetization, $M$, is normalized to its value at 0.4 T.

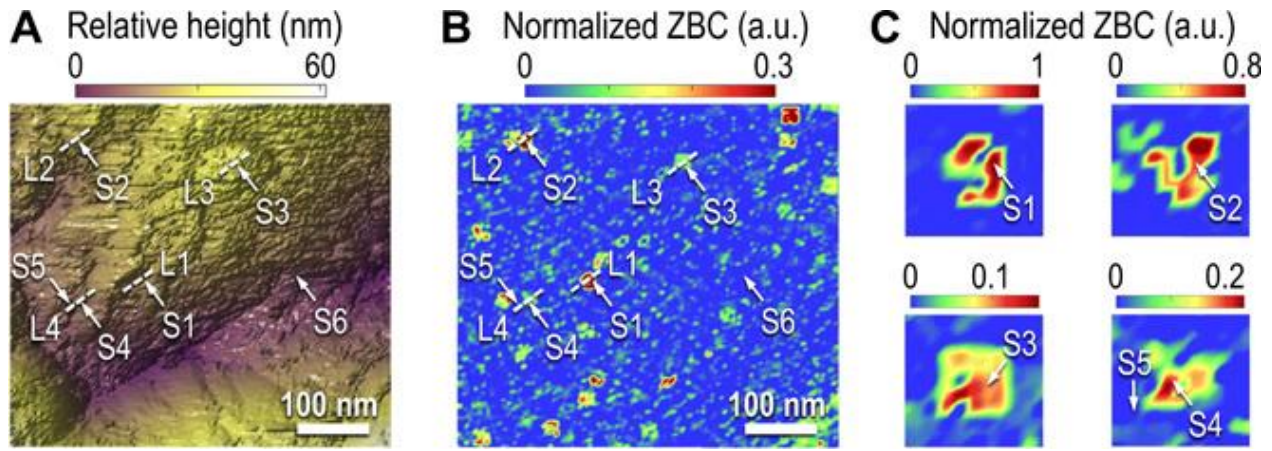

**Fig. 2. Topography and spectroscopic mapping of zero-bias conductance in hydrogenated boron-doped diamond.** (**A**) Topography of the diamond surface. Arrows (S1-S6) indicate the location of the characteristic Yu-Shiba-Rusinov bands as shown in Fig. 3A and 4A, and dashes (L1-L4) indicate the direction of the spectroscopic mapping in Fig. 3A. (**B**) Zero-bias conductance map acquired at 0.5 K, well below the superconducting transition temperature. The map was constructed by normalizing the tunneling spectra to the d$I$/d$V$ value at 5 mV, far outside the superconducting gap, and measuring the zero-bias conductance. The arrows and dashes indicate the same locations in (A), respectively. (**C**) Magnifications (39 nm × 39 nm) of the 'puddles' where the characteristic Yu-Shiba-Rusinov bands in Fig. 3 and 4 are observed.

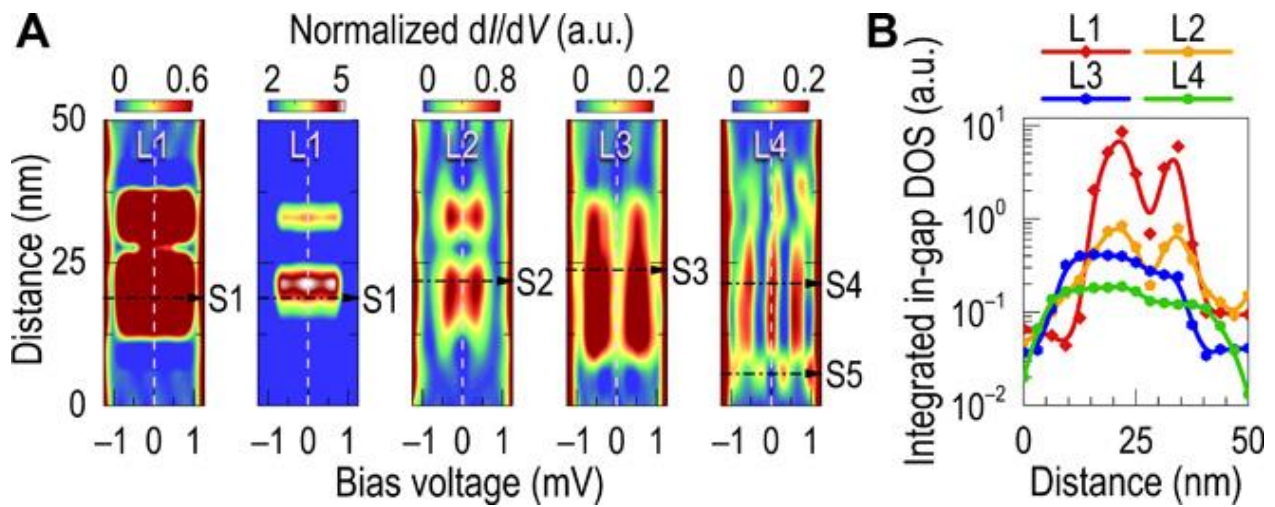

**Fig. 3. Yu-Shiba-Rusinov bands in hydrogenated boron-doped diamond.** (**A**) Spatial and energy evolutions of the tunneling conductance, d$I$/d$V$, along the dashes in Fig. 2. The tunneling conductance along L1 is shown in two panels with different color scales for the clarity of both spatial and energy evolutions. Dash-dotted arrows (S1-S5) indicate the location of the characteristic Yu-Shiba-Rusinov bands as shown in Fig. 4A. (**B**) Spatial evolution of the integrated in-gap density of states (DOS). The integration was performed over the range from -1 to +1 mV. The solid curves are a guide to the eye.

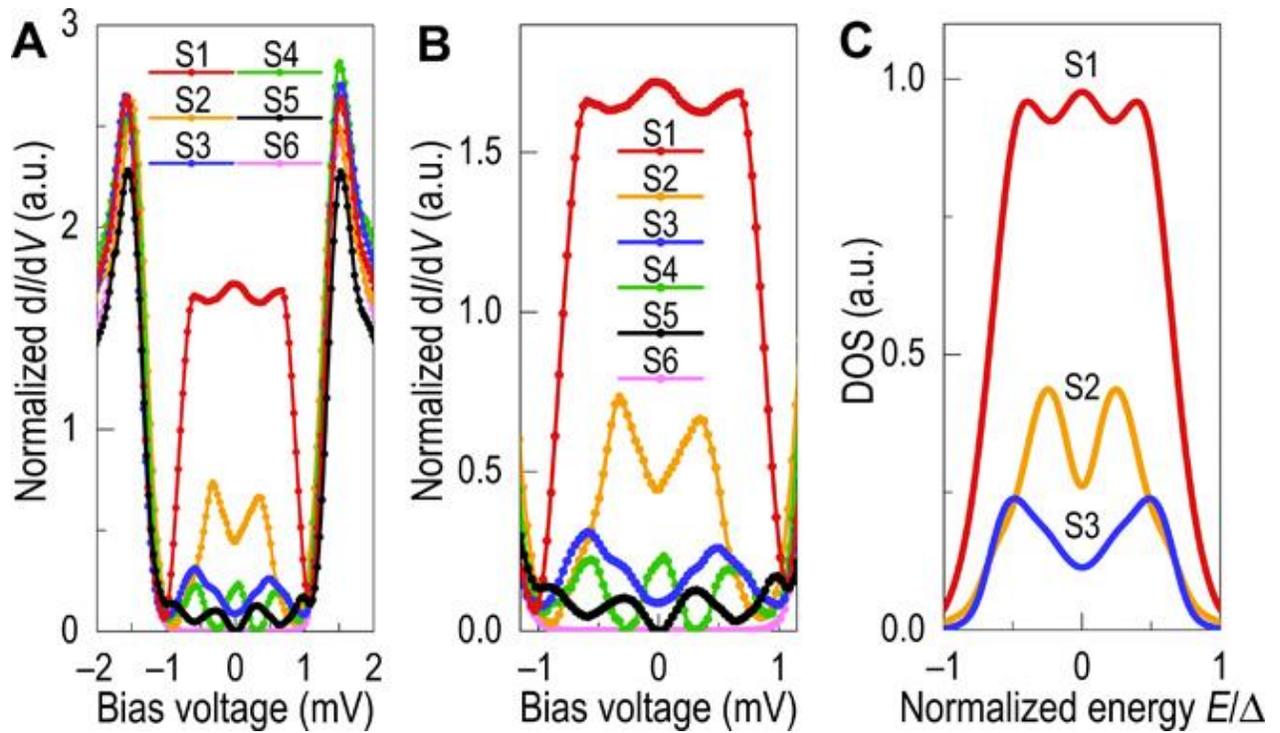

**Fig. 4. Comparison between measured and calculated Yu-Shiba-Rusinov bands.** (**A**) Characteristic experimental spectra acquired at S1-S6 in Fig. 2. The tunneling spectra are normalized to the conductance at 5 mV. (**B**) Magnification of the measured in-gap density of states. (**C**) Yu-Shiba-Rusinov bands calculated using tight-binding Bogoliubov-de Gennes equation in correspondence to the measured tunneling conductance spectra S1-S3.

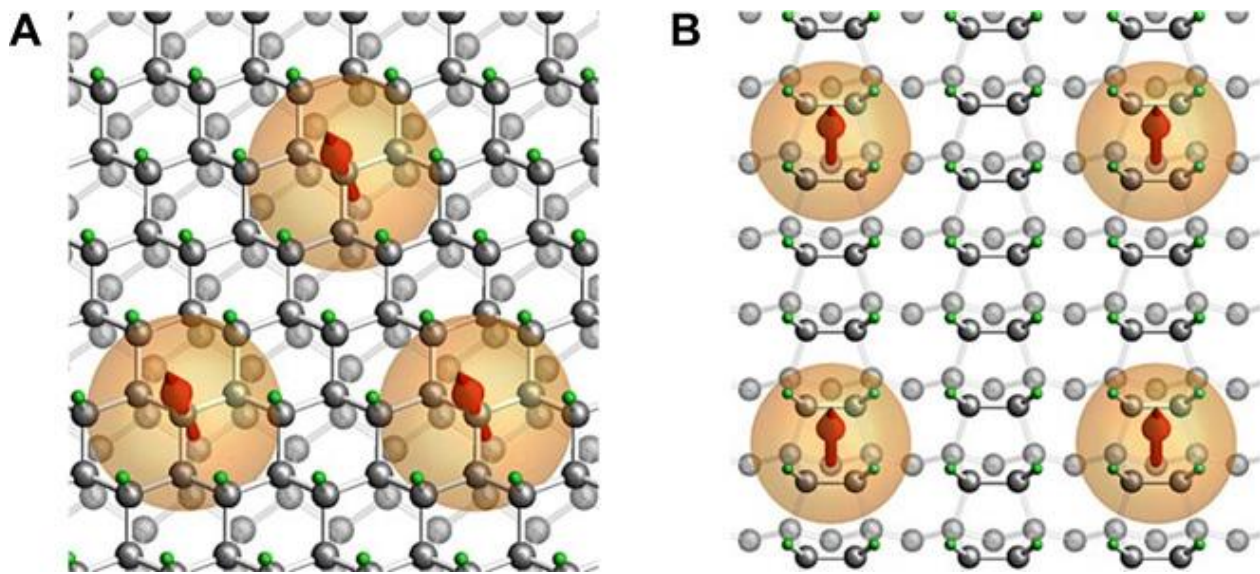

**Fig. 5. Arrangements of impurity spins at the diamond surface for tight-binding calculations of Yu-Shiba-Rusinov bands.** (**A**) Triangular spin lattice for the (111) surface. (**B**) Rectangular spin lattice for the (100) surface. Green and black pellets represent the surface hydrogen and carbon atoms, respectively. Red arrows and orange spheres denote the impurity spins and localized Yu-Shiba-Rusinov states, respectively. The hybridization of the localized Yu-Shiba-Rusinov states gives rise to the emergence of Yu-Shiba-Rusinov bands.